\documentclass[twocolumn,english,aps,manuscript,prl,reprint,longbibliography]{revtex4-1}

\usepackage[T1]{fontenc}
\usepackage[latin9]{inputenc}
\usepackage{amsmath}
\usepackage{amssymb}
\usepackage{graphicx}
\usepackage{natbib}
\usepackage{etoolbox}
\usepackage{bm}
\usepackage{subfigure}

\makeatletter

\usepackage[usenames,dvipsnames]{color}
\@ifundefined{textcolor}{}
{%
	\definecolor{BLACK}{gray}{0}
	\definecolor{WHITE}{gray}{1}
	\definecolor{RED}{rgb}{1,0,0}
	\definecolor{GREEN}{rgb}{0,1,0}
	\definecolor{BLUE}{rgb}{0,0,1}
	\definecolor{CYAN}{cmyk}{1,0,0,0}
	\definecolor{MAGENTA}{cmyk}{0,1,0,0}
	\definecolor{YELLOW}{cmyk}{0,0,1,0}
}
\definecolor{light-gray}{gray}{0.55}

\g@addto@macro\normalsize{%
	\setlength\abovedisplayskip{2pt}
	\setlength\belowdisplayskip{2pt}
	\setlength\abovedisplayshortskip{2pt}
	\setlength\belowdisplayshortskip{2pt}
}

\usepackage{babel}

\DeclareMathAlphabet{\mathpzc}{OT1}{pzc}{m}{it}
\makeatother

\begin{document}

\title{
 \LARGE Ergodic-localized junctions in a periodically-driven spin chain}


\author{Chen Zha$^{1,2}$}
\thanks{These authors contributed equally to this work.}
\author{V. M. Bastidas$^{3}$}
\thanks{These authors contributed equally to this work.}
\author{Ming Gong$^{1,2}$}
\thanks{These authors contributed equally to this work.}
\author{Yulin Wu$^{1,2}$}
\author{Hao Rong$^{1,2}$}
\author{Rui Yang$^{1,2}$}
\author{Yangsen Ye$^{1,2}$}
\author{Shaowei Li$^{1,2}$}
\author{Qingling Zhu$^{1,2}$}
\author{Shiyu Wang$^{1,2}$}
\author{Youwei Zhao$^{1,2}$}
\author{Futian Liang$^{1,2}$}
\author{Jin Lin$^{1,2}$}
\author{Yu Xu$^{1,2}$}
\author{Cheng-Zhi Peng$^{1,2}$}
\author{J. Schmiedmayer$^{4,5}$}
\author{Kae Nemoto$^{6}$}
\author{Hui Deng$^{1,2}$}
\author{W. J. Munro$^{3}$}
\email{bilmun@qis1.ex.nii.ac.jp}
\author{Xiaobo Zhu$^{1,2}$}
\email{xbzhu16@ustc.edu.cn}
\author{Jian-Wei Pan$^{1,2}$}
\affiliation{$^{1}$Hefei National Laboratory for Physical Sciences at Microscale and Department of Modern Physics, University of Science and Technology of China, Hefei, Anhui 230026, China}
\affiliation{$^{2}$Shanghai Branch,CAS Center for Excellence and Synergetic Innovation Center in Quantum Information and Quantum Physics, University of Science and Technology of China, Shanghai 201315, China.}

\affiliation{$^{3}$NTT Basic Research Laboratories \& Research Center for Theoretical Quantum Physics, 3-1 Morinosato-Wakamiya, Atsugi, Kanagawa, 243-0198, Japan}

\affiliation{$^{4}$Vienna  Center  for  Quantum  Science  and  Technology,Atominstitut,  TU  Wien,  Stadionallee  2,  1020  Vienna,  Austria}
\affiliation{$^{5}$Wolfgang  Pauli  Institute  c/o  Faculty  of  Mathematics,University  of  Vienna,  Oskar-Morgenstern  Platz  1,  1090  Vienna,  Austria}

\affiliation{$^{6}$National Institute of Informatics, 2-1-2 Hitotsubashi, Chiyoda-ku, Tokyo 101-8430, Japan}

\date{\today}


\maketitle
\noindent\textbf{We report the analogue simulation of an ergodic-localized junction by using an array of 12 coupled superconducting qubits. To perform the simulation, we fabricated a superconducting quantum processor  that is divided into two domains: a driven domain representing an ergodic system, while the second is localized under the effect of disorder. Due to the overlap between localized and delocalized states, for small disorder there is a proximity effect and localization is destroyed. To experimentally investigate this, we prepare a microwave excitation in the driven domain and explore how deep it can penetrate the disordered region by probing its dynamics. Furthermore, we performed an ensemble average over $50$ realizations of disorder, which clearly shows the proximity effect. Our work opens a new avenue to build quantum simulators of driven-disordered systems with applications in condensed matter physics and material science}.

Most of today's technology is based on our ability to manipulate quantum states of matter. At low temperatures, the collective motion of interacting particles lead to symmetry-broken phases such as ferromagnets and superconductors~\cite{Shahar1997}. Recently, there is an increasing interest in quantum phases of matter involving global properties of the energy spectrum that exist beyond this low-temperature limit~\cite{Altshuler2006, nandkishore15,Vosk2015,khemani17}. Among them, ergodic and localized phases not only play a fundamental role in our understanding of statistical mechanics, but they are of utmost importance in future quantum technologies. For example, ergodicity is intimately related to recent experiments of random sampling with superconducting qubits~\cite{Martinis2018,Martinis2019},  while  localized phases can be used as building blocks for quantum memories~\cite{Huse2014}.  When a many-body system is in its ergodic phase, interactions distribute the energy among all the available states and the expectation values of the local observables can be described by a Gibbs ensemble when energy is conserved~\cite{Srednicki1994, dalessio16}.  Contrary to this, in localized systems, disorder prevents the system from thermalizing due to the emergence of an extensive number of local conserved quantities~\cite{nandkishore15,Vosk2015,khemani17, chiaro2019}. Recent experiments in platforms ranging from superconducting qubits to cold atoms have shown signatures of  ergodic and localized phases in undriven systems~\cite{Neill2016,roushan17,Fan2018,Monroe2016,Bloch2015,Gross2016, chiaro2019}.  When the system is affected by an external driving force, however, conserved quantities such as energy are destroyed and the nature of the ergodic and localized phases can change dramatically~\cite{Polkovnikov2013,Moessner2015,Abanin2015,Moessner2016,Brandes2016,Refael2018}. A low-frequency drive can destroy local conserved quantities characteristic of a localized phase and bring the system to an infinite-temperature state~\cite{Rigol2014,Khemani2016}. Similarly, a high-frequency driving can destroy tunneling in many-body systems, leading to localization~\cite{Haenggi1998,Holthaus2005}. 

\begin{figure*}
 \centering
 \includegraphics[width=0.80\textwidth]{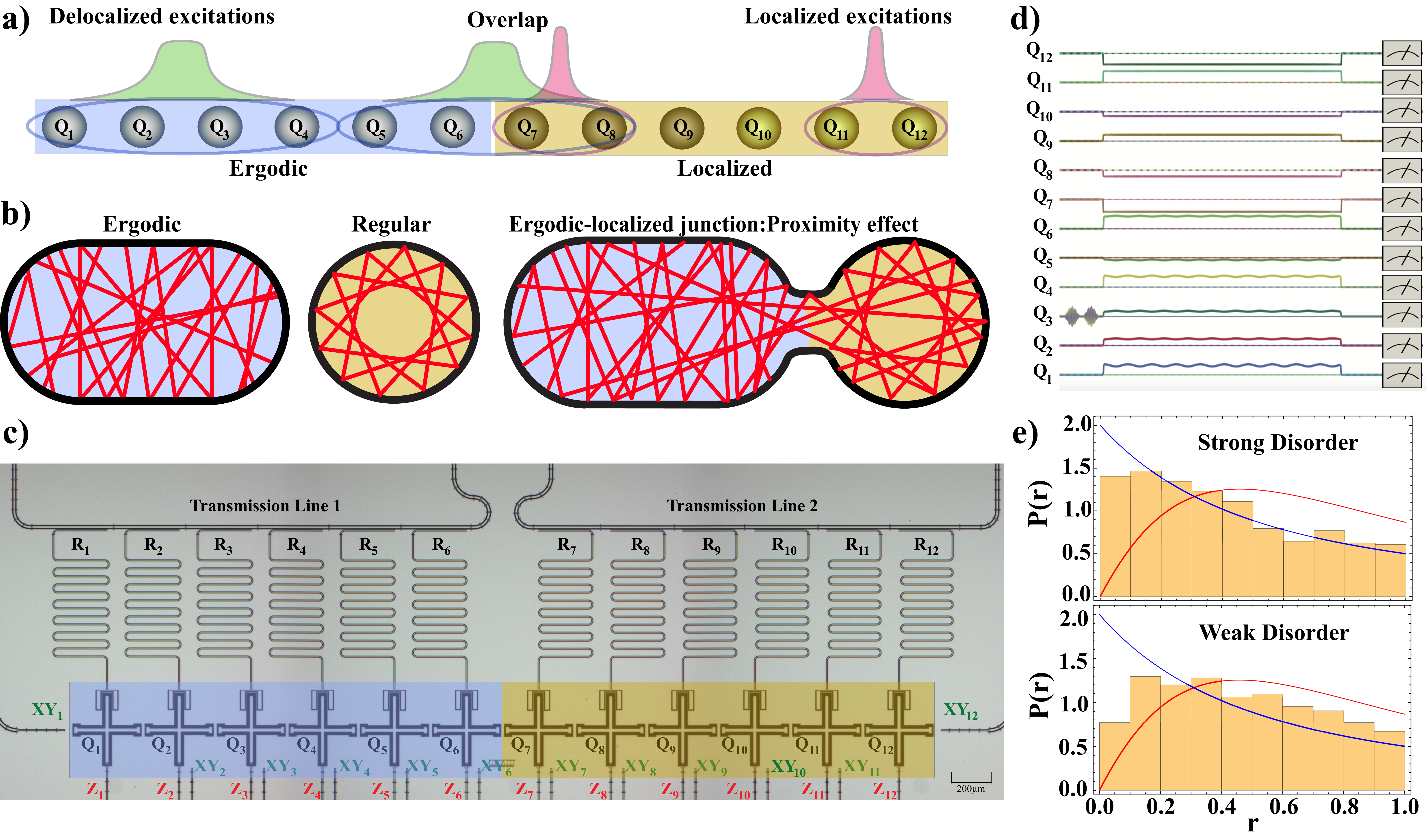}
	\caption{\textbf{Ergodic-localized junction with superconducting qubits.} \textbf{a)} When the driven domain is coupled to a disordered one, there is an overlap between localized and delocalized states. \textbf{b)} Depicts stadium and circular billiards, which exhibit ergodic and regular behavior, respectively. If one creates a junction between the two billiards, the destruction of conserved quantities creates a proximity effect, and the system explores all the available configurations. \textbf{c)} Optical micrograph of the superconducting chip. There are twelve Xmon variant of transmon qubits arranged in a 1D lattice. Each qubit has a
line for the XY control and a flux bias line for the Z control. In addition, each qubit is dispersively coupled to a resonator for state readout. \textbf{d)} Experimental waveform sequences to generate the ergodic-localized junctions. The ergodic domain is formed by qubits $Q_{1}-Q_{6}$, which are periodically driven and the localized domain is designed by adding disorder to qubits $Q_{7}-Q_{12}$. \textbf{e)} Shows the quasienergy level statistics~\cite{roushan17, Rigol2014, tangpanitanon2019quantum} of the ergodic-localized junction for an array of $L=12$
qubits for several realizations of disorder. For strong disorder strength $W=10J$, where $J$ is the coupling between qubits, the overlap between the two domains is small and the level statistics $P(r)$ is close to Poisson (blue line), indicating localization. As long as we decrease the disorder $W=3J$, the statistics become close to the circular orthogonal ensemble (red line), which is a signature of the proximity effect~\cite{Bastidas2018}. }
\label{Fig1}
\end{figure*}

%
Recent work has predicted that when disordered and driven domains are coupled, localization can be destroyed due to the overlap between localized and delocalized states~\cite{Bastidas2018},  as in Fig.~\ref{Fig1}~a). The localized domain becomes unstable and excitations can penetrate regions of the system that are otherwise forbidden. This induces a proximity effect: correlations are built up between the two domains and a portion of the localized domain becomes ergodic. The latter allows us to explore the stability of the localized phase when a disordered domain is coupled to a non-equilibrium environment. 

It is important to have an intuitive picture of the proximity effect in mind, so let us consider a particle moving in stadium and circular billiards, as illustrated in Fig.~\ref{Fig1}~b). In the circular billiard, the conservation of energy and angular momentum leads to regular behavior~\cite{dalessio16}. Ergodicity is broken and the system is only able to explore certain regions of phase space. In contrast, the stadium billiard does not preserve angular momentum and the motion becomes ergodic~\cite{dalessio16}. In this case, the system can explore all the available configurations in phase space with a constant energy. A natural question that arises is: what happens if one couples these two types of billiards?
As depicted in Fig.~\ref{Fig1}~b), to do this, one needs to break symmetries of the system and now a particle can explore regions that were forbidden by conservation rules. This can be interpreted as a proximity effect~\cite{Bastidas2018}.

The key issue is how do we explore these ergodic localized junctions. Superconducting quantum processors~\cite{roushan17,RoushanChiral2017, Martinis2018,chiaro2019, Martinis2019,Yan2019,Yangsen2019,2019Schuster} provide us with a high degree of controllability, which makes them a suitable platform to investigate the dynamics of driven systems. In this work, we experimentally simulate the proximity effect using a linear array of $N=12$ superconducting qubits, governed by the Hamiltonian
\begin{figure*}
 \centering
 \includegraphics[width=0.85\textwidth]{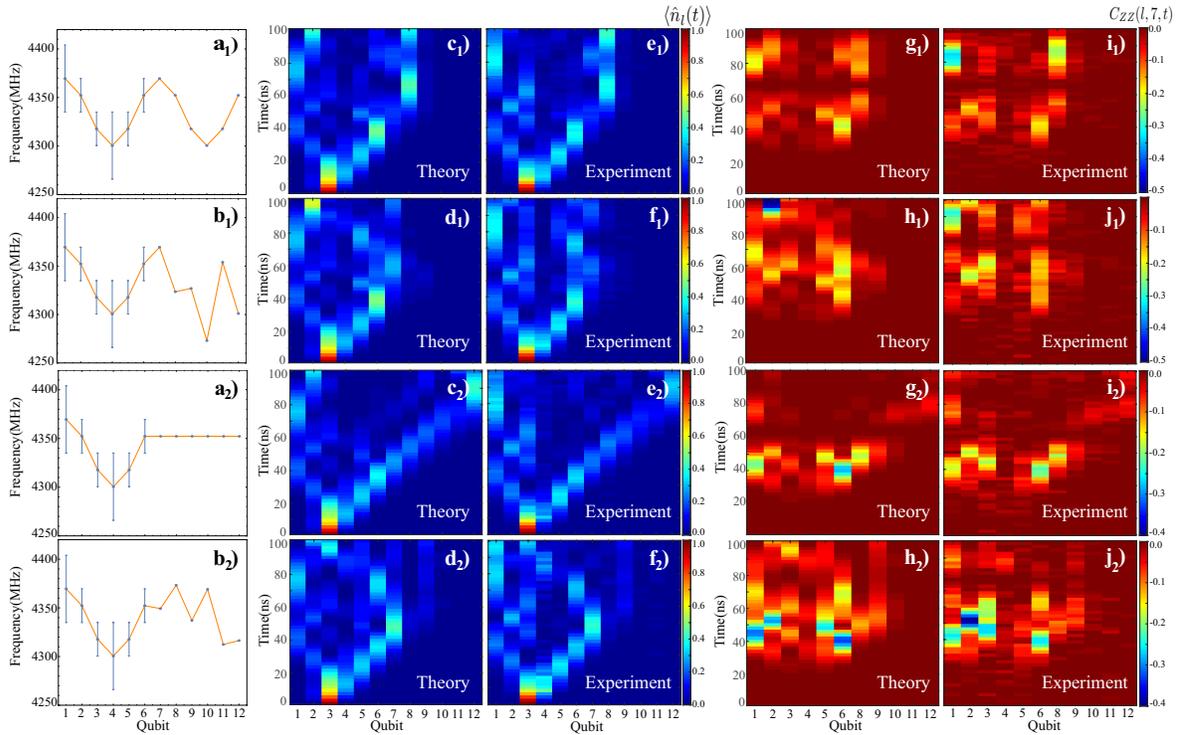}
	\caption{\textbf{Dynamics of the populations and correlations: Comparison between theory and experiment for  cosine and flat potentials of qubits $Q_7$ to $Q_{12}$.} To explore localization due to the competition between kinetic and potential energy, we consider a cosine potential for the qubits $Q_7$ to $Q_{12}$. Similarly, to investigate localization due to disorder, we consider a flat potential for the qubits $Q_7$ to $Q_{12}$.  The  panels \boldsymbol{$\textbf{a}_{1,2})$} to \boldsymbol{$\textbf{i}_{1,2})$} and \boldsymbol{$\textbf{b}_{1,2})$} to \boldsymbol{$\textbf{j}_{1,2})$} depict the results without  $(W=0)$ and with disorder $(W=5J)$, respectively.                                                                                                                                                                                       \boldsymbol{$\textbf{a}_{1,2})$} and \boldsymbol{$\textbf{b}_{1,2})$} show the qubit frequency setups used in the experiment. \boldsymbol{$\textbf{c}_{1,2})$} to \boldsymbol{$\textbf{f}_{1,2})$} depict the population dynamics $\langle\hat{n}_l\rangle$. Correspondingly, the panels \boldsymbol{$\textbf{g}_{1,2})$} to \boldsymbol{$\textbf{j}_{1,2})$} show the dynamics of the correlation function $C_{ZZ}(l,7,t)$. Due to the shape of the cosine potential, the excitation can penetrate a small region of the disordered domain, even in the absence of disorder. In contrast, the excitation can propagate ballistically for a flat potential in the absence of disorder.
}
\label{Fig2}
\end{figure*}
\begin{footnotesize}
\begin{align}
\label{eq:HamiltonianResonatorArray}
\hat{H}(t)&=\hbar\sum^{N}_{l=1}\left[g_{l}(t) \hat{n}_{l}+\frac{U}{2} \hat{n}_{l}(\hat{n}_{l}-1)\right]+\hbar J\sum^{N-1}_{l=1}(\hat{a}^{\dagger}_{l}\hat{a}_{l+1}+\hat{a}_{l}\hat{a}^{\dagger}_{l+1})
\ ,
\end{align}
\end{footnotesize}

 where $\hat{n}_{l}=\hat{a}^{\dagger}_{l}\hat{a}_{l}$ is the number operator, while $\hat{a}_{l}$ and $\hat{a}^{\dagger}_{l}$ are the bosonic annihilation and creation operators of an excitation at site $l$, respectively. Figure~\ref{Fig1}~c) depicts an optical micrograph of the chip designed for our experiment.
The qubit frequencies $g_{l}(t)/2\pi$ can be tuned with DC and pulse signals through the Z control lines. An example of the pulse sequences used during the experiment is shown in Fig.~\ref{Fig1}~d). The nonlinearity $U\approx-250~\text{MHz}\approx-22J$ is the strength of the onsite interactions while $J/2\pi=11.5$ MHz is the hopping strength between nearest-neighbor sites (see supp. material~\cite{SupplementalInfo} for further details).

To perform our quantum simulation of the ergodic-localized junction, we decompose the Hamiltonian~\eqref{eq:HamiltonianResonatorArray} in three pieces: $\hat{H}(t)=\hat{H}_{\text{erg}}(t)+\hat{H}_{\text{loc}}+\hat{H}_{\text{int}}$, where $\hat{H}_{\text{erg}}(t)$ and $\hat{H}_{\text{loc}}$ are the Hamiltonians of the ergodic and localized domains, respectively, while $\hat{H}_{\text{int}}$ is the interaction between them.  For a single excitation propagating through the array, the onsite energies and the hopping play the role of potential and kinetic energies, respectively.
In this case the Hamiltonian of the ergodic domain with qubits $Q_1-Q_{6}$ has the form
\begin{align}
\label{eq:HamiltonianErgodic}
\hat{H}_{\text{erg}}(t)&=\hbar\sum^{6}_{l=1}g_{l}(t) \hat{n}_{l}+ \hbar J\sum^{5}_{l=1}(\hat{a}^{\dagger}_{l}\hat{a}_{l+1}+\hat{a}_{l}\hat{a}^{\dagger}_{l+1}) \ ,
 \end{align}
where  $g_{l}(t)=\bar{g}+[\Delta_0+\Delta_1\cos(\omega t)]\cos(4\pi l/N)$ are modulated in space and time.  Here $\bar{g}/2\pi$ is the frequency of the rotating frame while $\Delta_0$ and $\Delta_1$ are the DC and AC components of the external drive, respectively.  In the experimental setup, we tuned the drive such that the resonance condition $3\omega=2\Omega$ is fulfilled with $\Delta_0=\Delta_1=3J$, where $\Omega=4\pi\sqrt{2\Delta_0 J}/N$ is the frequency of the small oscillations (see supp. material~\cite{SupplementalInfo} for further details). To achieve finite-size signatures of parametric resonance~\cite{Bastidas2010}, we drive with a frequency $\omega/2\pi=19.67$~MHz. In the classical model with $N\gg1$, stable fixed points become unstable and as we increase the amplitude of the drive, the dynamics become chaotic. In the finite system, however, quantum signatures of the chaotic motion appear in the level spacings distribution of the quasienergies~\cite{Rigol2014}, as we explain below. Next the localized domain with qubits $Q_{7}-Q_{12}$ is governed by 
\begin{align}
\label{eq:HamiltonianLocalized}
\hat{H}_{\text{loc}}&=\hbar\sum^{12}_{l=7}g_{l} \hat{n}_{l}+ \hbar J\sum^{11}_{l=7}(\hat{a}^{\dagger}_{l}\hat{a}_{l+1}+\hat{a}_{l}\hat{a}^{\dagger}_{l+1}) \ ,
 \end{align}
where we consider a spatial modulation of the qubit angular frequencies with cosine $g_{l}=\bar{g}+G_l+\Delta_0\cos(4\pi l/N)$ and flat $g_{l}=\bar{g}+G_l+\Delta_0$  background potentials. The disorder $G_l \in [-W,W]$ is drawn from a random uniform distribution with strength $W$. Finally, we consider a coupling between the ergodic and localized domains represented by the Hamiltonian $\hat{H}_{\text{int}}=\hbar J(\hat{a}^{\dagger}_{6}\hat{a}_{7}+\hat{a}_{6}\hat{a}^{\dagger}_{7})$.

\begin{figure*}
 \centering
 \includegraphics[width=0.85\textwidth]{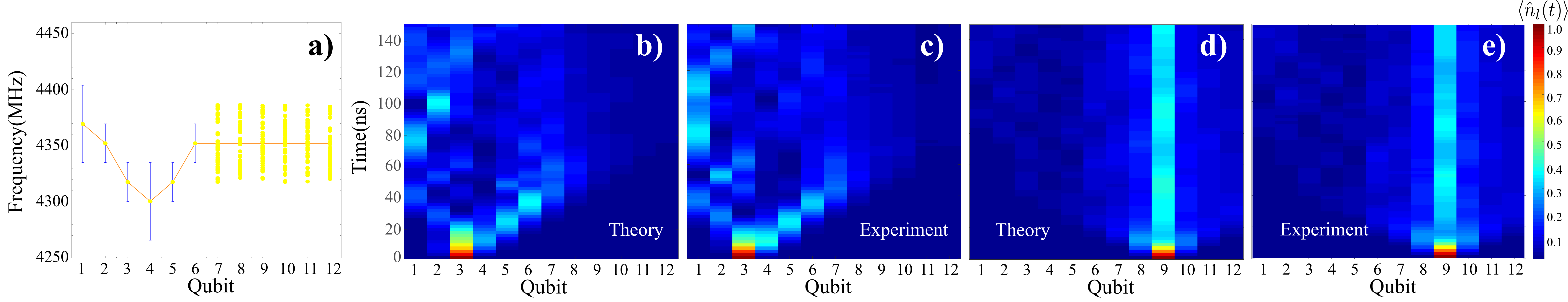}
	\caption{\textbf{Ensemble average of population dynamics.}  Motivated by the discussion on spectral signatures of the proximity effect, we explore the dynamics of the ensemble averages for a microwave excitations initially prepared at sites $l=3$ and $l=9$ in the ergodic and localized domains, respectively. \textbf{a)} Depicts the frequency setup for several realizations of disorder with strength $W=3J$. In average, the potential of qubits $Q_7$ to $Q_{12}$ is flat. The panels \textbf{b)}, \textbf{d)  show} the theoretical prediction for the averaged populations and {panels \textbf{c)}, \textbf{e)} } the experimental result. Although the dynamics for the individual instances can be quite different due to the local features of the potential, the average clearly shows the proximity effect. 
 }
\label{Fig4}
\end{figure*}

The proximity effect appears as a consequence of the overlap between delocalized and localized states in the ergodic and disordered domains, respectively. Therefore, to probe it, we can measure the dynamics of the populations to {determine} to which extent an excitation can penetrate the disordered domain. We start the experiment with the qubits in their idle frequencies and apply a local X gate at site $l=3$ to create the excitation. {In the absence of disorder, the drive induces transitions between site $l=3$ and the qubits $Q_7-Q_{12}$, which allows the excitation to penetrate this region [see Fig.~\ref{Fig2}~$\text{c}_{1,2}$)].}

The Z lines allows us to quickly bring the qubits from the idle to the working point frequencies, while keeping the excitation localized. To generate an external drive in the ergodic domain, we apply an AC pulse to qubits $Q_1-Q_{6}$ by using the Z control lines. The disordered domain is engineered by designing a profile of random working frequencies for the qubits  $Q_7-Q_{12}$. 
We chose different modulations of the potential in order to study two different aspects of localization and their influence on the proximity effect. A cosine background potential enables us to study localization due to the potential shape, while with a flat background potential we explore localization due to disorder.  
The comparison between the theoretical prediction and the experimental results for the population dynamics is shown in Figure~\ref{Fig2}. The rows {$\text{a}_{1,2})$ to $\text{i}_{1,2})$ and $\text{b}_{1,2})$ to $\text{j}_{1,2})$} depict the dynamics without $(W=0)$ and with disorder $(W=5J)$, respectively. {Figures~\ref{Fig2}~$\text{a}_{1,2})$~and~$\text{b}_{1,2})$} depict the working frequencies for the qubits. {The panels~$\text{c}_{1,2})$~and~$\text{d}_{1,2})$} show the theoretical prediction {and~$\text{e}_{1,2})$~and~$\text{f}_{1,2})$}  the experimental measurements of the populations $\langle\hat{n}_{l}(t)\rangle$ at site $l$ as a function of time. Our experimental setup also allows us to access the full correlation matrix of the photons. The correlation function  $C_{ZZ}(l,7,t)=\langle\sigma^z_{l}(t)\sigma^z_{7}(t)\rangle-\langle\sigma^z_{l}(t)\rangle\langle\sigma^z_{7}(t)\rangle$ with $\sigma^z_{l}=2\hat{n}_{l}-1$ reveals signatures of the proximity effect {because it gives a measure of the overlap between localized and delocalized states. Panels~$\text{g}_{1,2})$,~$\text{h}_{1,2})$~and~$\text{i}_{1,2})$, $\text{j}_{1,2})$ show the theoretical and experimental dynamics of the correlation function, respectively}.

 In the case of a cosine potential, even in the absence of disorder, the states become localized once the kinetic energy $\hbar J$ is smaller than the potential energy $\hbar\Delta_0$. If we add disorder with strength $W\gg \Delta_0$~\cite{Glaetzle2017}, localization is enhanced because Anderson localization dominates.  On the contrary, in the case of a flat potential, localization is solely caused by disorder~\cite{anderson58,kramer93}. {In the absence of drive, a localized system exhibit is a set of local conserved quantities, which are given by the energies.} As a consequence, if we couple the localized domain to the ergodic one, these conserved quantities can be destroyed, as in the case of the billiards shown in Fig.~\ref{Fig1}~b).
In the absence of disorder, an initial excitation at site $l=3$ can penetrate up to site $l=8$, as one can see in {Figs.~\ref{Fig2}~$\text{c}_{1})$~and~$\text{e}_{1})$}. The disorder inhibits the tunneling to the localized domain as shown in {Figs.~\ref{Fig2}~$\text{d}_{1})$~and~$\text{f}_{1})$}. Due to the shape of the potential, the drive is not resonant with states in the disordered domain and therefore, the particle cannot penetrate the localized domain. In the case of a flat potential, the excitation propagates ballistically in the absence of disorder  as in {Figs.~\ref{Fig2}~$\text{c}_{2})$~and~$\text{e}_{2})$}. Disorder restrict the distance that the particle can travel within the disordered domain as shown in {Figs.~\ref{Fig2}~$\text{d}_{2})$~and~$\text{f}_{2})$}. For both cosine and flat potentials, the dynamics of the correlations in Fig.~\ref{Fig2} also show signatures of the proximity effect.  The latter creates an overlap between delocalized and localized states, which induces correlations between sites in different domains.
Thus regions in the disordered domain that cannot be accessed by the microwave excitation remain uncorrelated from the driven domain.

So far we have discussed dynamical signatures of the proximity effect that appear in the populations $\langle\hat{n}_{l}(t)\rangle$ and the correlations $C_{ZZ}(l,7,t)$. Next, we theoretically explore spectral signatures of the proximity effect. In order to do this, we numerically construct the  {Floquet operator $\hat{\mathcal{F}}=\hat{U}(T)$, where $\hat{U}(T)$ is evolution operator} in a period $T=2\pi/\omega=50.84$~ns of the drive. {By calculating the eigenvalues $\exp(-\mathrm{i}\varepsilon_{\alpha}T/\hbar)$ of  $\hat{\mathcal{F}}$, we obtain the quasienergies $-\hbar\omega/2\leq\varepsilon_{\alpha}\leq\hbar\omega/2$ and the statistics of the ratios $r_{\alpha}=\min(\delta_{\alpha},\delta_{\alpha+1})/\max(\delta_{\alpha},\delta_{\alpha+1})$ between gaps $\delta_{\alpha}=\varepsilon_{\alpha+1}-\varepsilon_{\alpha}$}~\cite{roushan17, Rigol2014, tangpanitanon2019quantum}.
When the system is fully ergodic, there is a strong level repulsion and the states are delocalized in space. Thus, ratios $r_{\alpha}$ follow the same statistics as the circular orthogonal ensemble of random matrices. Contrary to this, when the states are localized, the energies are uncorrelated and the statistics follows a Poissonian distribution. Fig.~\ref{Fig1}~e) shows the statistics of levels for our qubit array. For convenience, we perform the statistical analysis in the case of an average flat potential of the localized domain. 
For strong disorder $W=10J$, the driving is not able to destroy localization in the system and the localized domain remains stable. Therefore, the statistics is close to a Poissonian distribution (blue curve). For weaker disorder $W=3J$, the distribution resembles the statistics of the circular ensemble (red curve). This is a signature of the proximity effect: the drive destroys integrals of motion {and the excitations can penetrate regions in the localized domain that were forbidden}.

In the case of a disordered potential, the dynamics of the populations can be quite different depending on the chosen instance of disorder. This is because the drive can induce resonances that appear randomly for each instance, leading to tunneling between resonant sites. For this reason, it is convenient to average the experimental results over several realizations. In our experimental setup, this can be easily achieved {by using the control Z lines of the superconducting quantum processor to define a profile of random working frequencies}. Fig.~\ref{Fig4}~a) shows the frequency setup for $50$ instances of disorder with strength $W=3J$. To build the ensemble average, we propagate an excitation initially prepared at site $l=3$ for each realization of disorder. After that, we calculate the ensemble average of the measured populations for times up to $t=150$~ns. The evolution time is limited by the relaxation $T_1$ and dephasing $T_2^*$ times of the qubit array. As we show in the supplemental information~\cite{SupplementalInfo}, the $Q_9$ has the shortest time scales of $T_1=35.13\mu$s and $T_2^*=1.79\mu$s. Figs.~\ref{Fig4}~b)~and~c) show the theory and the experiment, respectively. In contrast to the dynamics of the individual realizations of disorder, the ensemble average shows a clear signature of the proximity effect. In average, the drive destroys localization and the excitation can penetrate the disordered domain.  {Similarly, Figs.~\ref{Fig4}~d)~and~e) show that an excitation initially prepared at site $l=9$ has a small probability to travel the ergodic domain. In the absence of the drive, the tunneling to the domain with qubits $Q_1-Q_6$ is suppressed due to Anderson localization.}

In summary, we have simulated the proximity effect by using a superconducting quantum processor and demonstrated that its signatures appear in the populations and correlations at the level of a single realization of disorder. We go beyond this, and experimentally explore the dynamics of the disorder average over 50 realizations. This shows clear signatures of how a particle in the ergodic domain can penetrate the localized domain due to the proximity effect.  
Our work will inspire future experiments in platforms such as cold atoms, Rydberg atoms and trapped ions.
We envision potential applications of our results to quantum memories, as one can in principle store information in the localized domain. As long as it is stable, the information is robust against the coupling to an ergodic domain. The local control and addressability of the microwave excitations achieved with our platform will open a new avenue to explore the stability of other phases of matter under the effect of quantum baths out of equilibrium.

\textit{Acknowledgement:\textemdash}
We thank M. P. Estarellas, I. Mahboob, and A. Sakurai for valuable discussions. 
The authors thank the USTC Center for Micro- and Nanoscale Research and Fabrication for supporting the sample fabrication. The authors also thank Shujuan Li for her help in sample fabrication. The authors also thank QuantumCTek Co., Ltd., for supporting the fabrication and the maintenance of room-temperature electronics. This research was supported by the National Basic Research Programme (973) of China (Grant no. 2017YFA0304300), the Chinese Academy of Sciences, Anhui Initiative in Quantum Information Technologies, Technology Committee of Shanghai Municipality, National Natural Science Foundation of China (Grants no. 11574380 and {no. 11905217}) and Natural Science Foundation of Shanghai (Grant No. 19ZR1462700). 
This work was supported in part from the Japanese program Q-LEAP, the MEXT KAKENHI Grant-in-Aid for Scientific Research on Innovative Areas Science of hybrid quantum systems Grant No.15H05870 and the JSPS KAKENHI Grant No. 19H00662. This project was also made possible through the support of a grant from the John Templeton Foundation (JTF 60478).

\end{document}


\title{Supplementary materials for ``Ergodic-localized junctions in a periodically-driven spin chain" 
}

\author{Chen Zha$^{1,2}$}
\thanks{These authors contributed equally to this work.}
\author{V. M. Bastidas$^{3}$}
\thanks{These authors contributed equally to this work.}
\author{Ming Gong$^{1,2}$}
\thanks{These authors contributed equally to this work.}
\author{Yulin Wu$^{1,2}$}
\author{Hao Rong$^{1,2}$}
\author{Rui Yang$^{1,2}$}
\author{Yangsen Ye$^{1,2}$}
\author{Shaowei Li$^{1,2}$}
\author{Qingling Zhu$^{1,2}$}
\author{Shiyu Wang$^{1,2}$}
\author{Youwei Zhao$^{1,2}$}
\author{Futian Liang$^{1,2}$}
\author{Jin Lin$^{1,2}$}
\author{Yu Xu$^{1,2}$}
\author{Cheng-Zhi Peng$^{1,2}$}
\author{J. Schmiedmayer$^{4,5}$}
\author{Kae Nemoto$^{6}$}
\author{Hui Deng$^{1,2}$}
\author{W. J. Munro$^{3}$}
\email{bilmun@qis1.ex.nii.ac.jp}
\author{Xiaobo Zhu$^{1,2}$}
\email{xbzhu16@ustc.edu.cn}
\author{Jian-Wei Pan$^{1,2}$}
\affiliation{$^{1}$Hefei National Laboratory for Physical Sciences at Microscale and Department of Modern Physics, University of Science and Technology of China, Hefei, Anhui 230026, China}
\affiliation{$^{2}$Shanghai Branch,CAS Center for Excellence and Synergetic Innovation Center in Quantum Information and Quantum Physics, University of Science and Technology of China, Shanghai 201315, China.}

\affiliation{$^{3}$NTT Basic Research Laboratories \& Research Center for Theoretical Quantum Physics, 3-1 Morinosato-Wakamiya, Atsugi, Kanagawa, 243-0198, Japan}

\affiliation{$^{4}$Vienna  Center  for  Quantum  Science  and  Technology,Atominstitut,  TU  Wien,  Stadionallee  2,  1020  Vienna,  Austria}
\affiliation{$^{5}$Wolfgang  Pauli  Institute  c/o  Faculty  of  Mathematics,University  of  Vienna,  Oskar-Morgenstern  Platz  1,  1090  Vienna,  Austria}

\affiliation{$^{6}$National Institute of Informatics, 2-1-2 Hitotsubashi, Chiyoda-ku, Tokyo 101-8430, Japan}


\let\oldthebibliography=\thebibliography
\let\oldendthebibliography=\endthebibliography
\renewenvironment{thebibliography}[1]{%
	\oldthebibliography{#1}%
	\setcounter{enumiv}{5}%
}{\oldendthebibliography}

\maketitle
\textbf{}
\vspace{-20pt}
\tableofcontents

\vspace{-20pt}

\renewcommand{\theequation}{S\arabic{equation}}
\setcounter{equation}{0}  

\renewcommand{\thefigure}{S\arabic{figure}}
\setcounter{figure}{0}  

\renewcommand{\thetable}{S\arabic{table}}
\setcounter{table}{0}

\section{Experimental device}

\subsection{Performance}

We perform our experiment on a twelve-qubit superconducting quantum processor mounted in a dilution refrigerator system whose wiring set-up is the same as reported in Ref~\cite{ye2019}. As shown in Fig.~1~c) in the main text, the array is composed by a one-dimensional array of $12$  transmon qubits~\cite{You2007,Koch2007,Barends2014}. Each Qubit capacitively couples to its nearest neighbors. The coupling strength between the qubits is designed to be $J/2\pi=11.5$~MHz, but the experimental measurements reveal a small deviation of this value for each pair of qubits, as listed in Table~\ref{tab1}. Furthermore, our measurements reveal that the couplings do not change too much as a functions of the qubit's frequencies~\cite{Yan2019}.

The qubit frequencies can be tuned by using DC and pulse signals through Z control lines. The DC signals allow us to set the qubits to their idle frequencies (Table \ref{tab1}). This allows us to obtain a better coherence, reduce the crosstalk between two qubits, and lower the interaction among the adjacent qubits. The energy relaxation time $T_1$ and the dephasing time $T_2^*$ of the qubits listed in Table \ref{tab1} are measured at their idle frequencies. The pulse signals enable the controlability of the effective coupling strength between the qubits.  By tuning the qubits from their idle frequencies to the working frequencies, we let the system evolve to perform the analogue quantum simulation. Furthermore, each qubit at site $l$ is dispersively coupled to its exclusive resonator $R_l$ for the state readout. The twelve resonators $R_1$ to $R_{12}$ are divided into two groups as $R1-R_{6}$ and $R_7-R_{12}$. Each group of resonators inductively couples to a transmission line, through which the probe signal allows one to measure the state of the corresponding qubits simutaneously. The probe signals are amplified by two corresponding impedance matching parametric amplifiers(IMPA)~\cite{Mutus2014} with amplification of 14.22~dB and 14.06~dB, respectively. The frequencies of the readout resonators are shown in Table~\ref{tab1}.

The idle frequencies of the $12$ qubits are optimized based on their coherence performance.
In addition, we choose the working frequencies of the two groups of qubits, i.e.,  $Q_1-Q_6$ and $Q_7-Q_{12}$, 
in order to define two different energy profiles. The first group of qubits $Q_1-Q_6$ has a frequency profile that approximately follows the formula $g^{\textrm{cosine}}_{l}=\bar{g}+\Delta_0\cos(4\pi l/N)$, where $l$ is the position along the chain, $\Delta_0$ is the amplitude of the spatial modulation and $\bar{g}$ is the  frequency of the rotating frame. As we need to explore two different aspects of localization in the domain $Q_7-Q_{12}$, in our experiment, we consider two different frequency profiles for this group. The first one is a frequency profile with a cosine dependence $g^{\textrm{cosine}}_{l}=\bar{g}+\Delta_0\cos(4\pi l/N)$ without disorder. The second one defines a flat potential $g_{l}=\bar{g}+G_l+\Delta_0$ under the effect of disorder $G_l \in [-W,W]$, which is drawn from a random uniform distribution with strength $W$. In our design, we take into account the limitations of the maximum amplitude of the pulse signals. The working frequencies in the absence of disorder are listed in Table \ref{tab1}. The strength of the disorder  for the second group of qubits $Q_7-Q_{12}$ is chosen to be $W=5J$.

        \begin{table*}[t]
	
		\scriptsize
		\renewcommand\arraystretch{1.6}
		\resizebox{\textwidth}{!}{
			\begin{tabular}{c |c c c c c c c c c c c c }
				\hline
				\hline
				Parameters
				&Q$_1$  &Q$_2$  &Q$_3$   &Q$_4$  &Q$_5$  &Q$_6$ &Q$_7$  &Q$_8$ &Q$_9$  &Q$_{10}$  &Q$_{11}$  &Q$_{12}$\\
				\hline
				$\omega_{\textrm{read}}/2\pi$  (GHz)
				&6.375	&6.430	&6.486	&6.521	&6.571	&6.634	&6.687	&6.723	&6.782	&6.835	&6.891	&6.946\\
				$\omega_{\textrm{max}}/2\pi$  (GHz)
				&4.578	&4.572	&4.707	&4.689	&4.614	&4.666	&4.380	&4.485	&4.702	&4.551	&4.469	&4.593\\
				$\omega_{\textrm{idle}}/2\pi $ (GHz)
				&3.975	&4.505	&4.070	&4.548	&4.030	&4.656	&3.964	&4.465	&4.004	&4.505	&4.060	&4.570\\
				$g^{\textrm{cosine}}_{l}/2\pi$  (GHz)
				&4.370 	&4.352 	&4.318 	&4.300 	&4.318 	&4.352 	&4.370 	&4.352 	&4.318 	&4.300 	&4.318 	&4.352\\
				$g^{\textrm{flat}}_{l}/2\pi$  (GHz)
				&4.370 	&4.352 	&4.318 	&4.300 	&4.318 	&4.352 	&4.352	&4.352 	&4.352 	&4.352 	&4.352 	&4.352\\				
				$T_1$ ($\mu$s)
				&55.32	&37.95	&28.19	&39.96	&31.85	&29.98	&74.40	&47.80	&35.13	&32.91	&58.54	&41.30\\
				$T_2^*$ ($\mu$s)
				&1.95	&4.86	&1.82	&2.37	&1.91	&9.24	&2.33	&6.26	&1.79	&3.63	&2.59	&11.57\\

				$J/2\pi $(MHz) &\multicolumn{12}{c}{\fontsize{10pt}{0}~~~11.37~~11.73~~11.87~~11.71~~11.47~~11.33~~11.33~~11.29~~11.24~~11.64~~12.00~~~~} \\
				$\eta/2\pi $ (MHz)
				&-248	&-256	&-264	&-216	&-256	&-248	&-240	&-256	&-256	&-248	&-248	&-240\\
	
				$\chi_{qr}/2\pi$ (MHz)
				&0.13	&0.18	&0.13	&0.16	&0.10	&0.19	&0.11	&0.15	&0.11	&0.25	&0.11	&0.15\\  	
				$f_{00}$
				&0.88	&0.92	&0.87	&0.92	&0.87	&0.97	&0.92	&0.91	&0.94	&0.97	&0.93	&0.98\\
				$f_{11}$
				&0.85	&0.83	&0.83	&0.87	&0.83	&0.91	&0.83	&0.86	&0.86	&0.89	&0.84	&0.86\\
				Readout visibility  	
				&0.73	&0.75	&0.70	&0.79	&0.69	&0.88	&0.76	&0.76	&0.80	&0.87	&0.77	&0.83\\
				Integration time (ns)
				&3900	&3500	&3500	&3000	&3500	&2700	&3100	&2500	&3200	&2800	&3400	&2800\\   	   	
				\hline
				\hline
			\end{tabular}

		}

		\caption{
			Device parameters: 
			$\omega_{\textrm{read}}/2\pi$ is the frequency of the readout resonator;
			$\omega_{\textrm{max}}/2\pi$ is the maximum frequency of the qubit; 
			$\omega_{\textrm{idle}}/2\pi$ is the qubit's idle frequency; 
			$T_1$ is the qubit's energy relaxation time; 
			$T_2^*$ is the qubit's dephasing time determined from Ramsey measurement; 
			$J$ is the nearest-neighbor qubits' coupling strength, of which values are listed between the neighboring qubits. 
			The on-site nonlinear interaction $U$  approximates the qubit's  anharmonicity ($\eta \equiv f_{12}-f_{01}$, with $f_{12}$ being the transition frequency between $|1\rangle$ and $|2\rangle$ and $f_{01}$ being the transition frequency between $|0\rangle$ and $|1\rangle$) measured near the idle points, and can be considered as a fixed value since it is almost constant with the qubit frequency; 
			$\chi_{qr}$ is the dispersive shift; 
			$f_{00}$ ($f_{11}$) is the probability of correctly reading out the qubit state when it is prepared in $|0\rangle$ ($|1\rangle$);
			the readout fidelity is defined as $f_{00}+f_{11}-1$. 
			The integration time for measurements of each qubit ranges from 2500~ns to 3900~ns, which results from the optimization of the readout visibility.}
		\label{tab1}	
		\end{table*}

\subsection{Fabrication }
The fabrication of our device was carried out by following the steps described below~\cite{ye2019}:%
\begin{itemize}
\item[(1)]  A 100~nm aluminum thin film is deposited on a 2~inch c-plane degassed sapphire wafer in the electron-beam evaporation system Plassys MEB 550SL3.
\item[(2)]  Laser lithography followed by wet etching are applied to defined and construct the transmission lines, control lines, readout resonators and capacitors
\item[(3)]  The golden alignment marks for electron-beam lithography are fabricated with laser lithography, thermal evaporation and liftoff.
\item[(4)]  Laser lithography and thermal evaporation of 200~nm $\text{SiO}_{2}$ are used to fabricate the insulation layer of the crossover.
\item[(5)]  The bridge of the crossover is built with laser lithography, electron-beam evaporation of 350~nm aluminum film and liftoff.
\item[(6)]  The wafer is diced into 10~mm by 10~mm chips.
\item[(7)]  Electron-beam lithography followed by double-angle evaporation and controlled oxidation in Plassys MEB 550SL3 are applied to designed and fabricate delicate structures such as the Josephson junctions.
\item[(8)]  The wafer is  diced into 8~mm by 8~mm chips.
\item[(9)] The processor is wire bonded to a copper sample box.
\end{itemize}

\subsection{Waveform sequences}
Our setup provide us with a high degree of freedom in controlling of the qubits' frequencies, which is achieved through the Z control lines. In our system, we can apply two kinds of signals through Z control lines tuning the frequencies of the qubits. The first kind is a DC signal, which fixes the qubits' frequencies to their idle frequencies, and is kept constant during the experiment. The other kind is a pulse signal, which is used to quickly detune the qubits from their idle points to working frequencies. In addition, through XY control lines, we can control the state of the qubits and create microwave excitations at any site along the chain. The local control $Z_l$ and $XY_l$ lines at site $l$ are depicted in the optical micrograph of the device shown in Fig.~1~c) in the main text.

\begin{figure}
	\centering
	\includegraphics[width=0.65\textwidth]{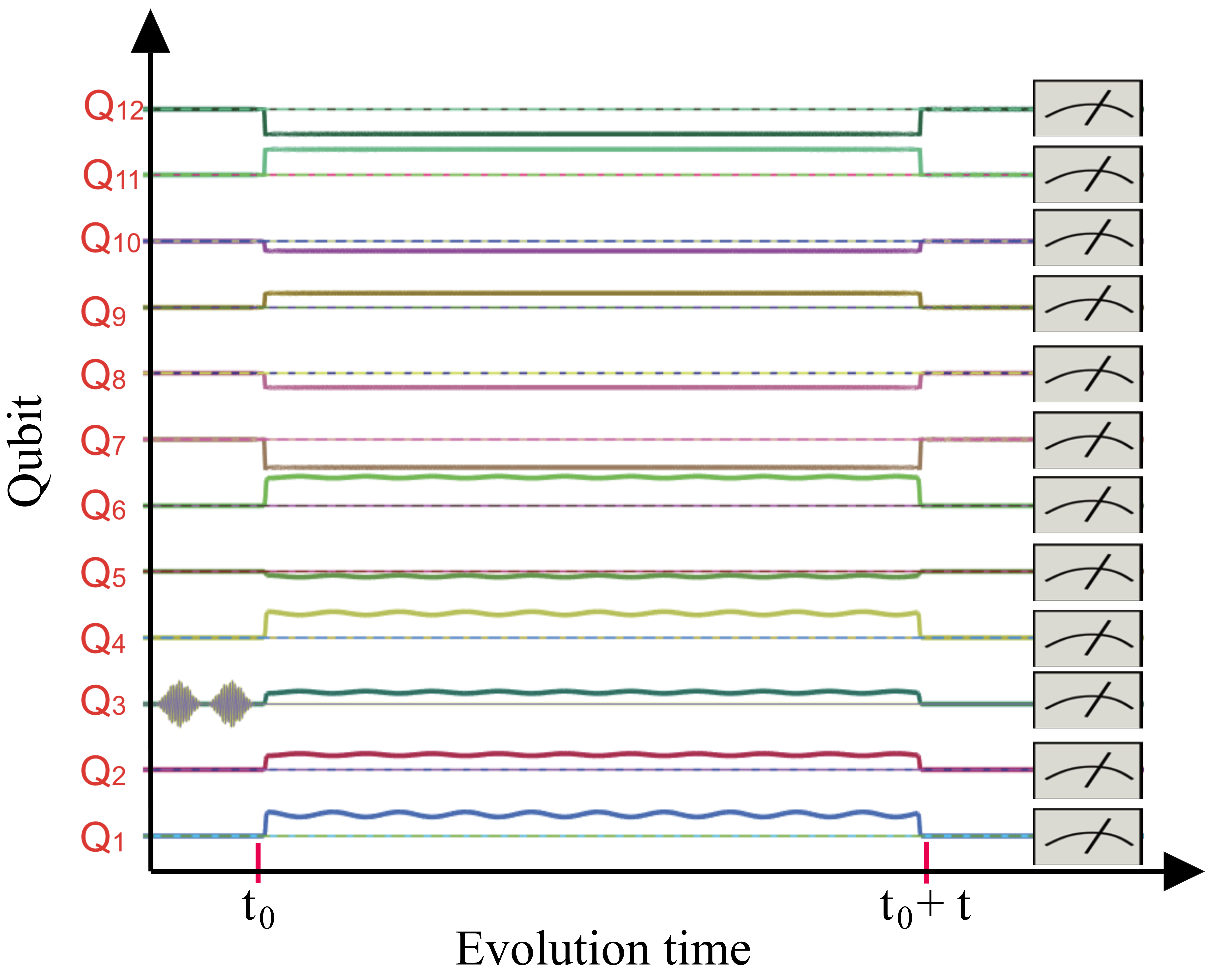}
	\caption{
		Experimental waveform sequences used during the experiment. By using the X control line, we create a microwave excitation at site $l=3$.  After this, all twelve qubits are quickly detuned to their working frequencies through Z control line. The group of qubits $Q_{1}-Q_{6}$ is used to simulate the ergodic domain. With this aim, we use the local Z control lines to drive the system. At the end of the evolution, all qubits are measured to obtain the expectation value $\langle\hat{a}_l^{\dagger}\hat{a}_l\rangle$, which gives us information about the populations.
	} 
	\label{fig2}
\end{figure}

The implementation of our experiment is divided into four steps: initialization, state preparation, designed evolution, and readout. In the following, we describe in detail the aforementioned steps:

\textit{Initialization:-} To initialize the processor, we wait for 200~$\mu$s in order to let the qubits decay to their ground state $|0\rangle$. 

\textit{State preparation:-}After initializing the qubits to their ground state, we prepare the desired initial state of the qubit at the $l$-th site by using local XY control lines. In order to create a single microwave excitation localized at site $l$, we apply a $\pi$-rotation along the $X$-axis that excites the qubit at site $l$ from $|0\rangle_l$ to $|1\rangle_{l}$. Afterwards, we apply pulse signals through the Z control lines to quickly detune all qubits to their working frequencies at time $t_0$, that denotes the initial time at which we start the quantum simulation. 

\textit{Designed evolution:-} In our experimental setup, the qubits $Q_{1}-Q_{6}$ are used to simulate the ergodic domain. In order to simulate ergodic behavior in this group of qubits, we apply waveforms on Z lines to drive the system. This allows us to modulate the frequencies of the first group of qubits in the formula of space and time: $g_{l}(t)=\bar{g}+[\Delta_0+\Delta_1\cos(\omega t)]\cos(4\pi l/N)$. The frequency $\omega/2\pi=19.67$~MHz of the modulation is chosen such that it satisfies the resonance condition $m\omega=2\Omega$ with $m=3$, where $\Omega=4\pi\sqrt{2\Delta_0 J}/N$ is the frequency of the small oscillations, and $\Delta_0=\Delta_1=3J$. In contrast to this, the frequencies of the second group of qubits $Q_{7}-Q_{12}$ are kept constant during each experiment to simulate the localized domain. We design different profiles of random working frequencies to simulate the effect of disorder in this group of qubits.  After tuning the qubits from the idle points to their working frequencies at time $t_0$, we apply pulses through the control Z lines to perform the analogous simulation of the ergodic-localized junction. With this aim, we let the system evolve for a time interval $t_0<\tau<t_0+t$, where $t=150\text{ns}$. The pulse sequences used in our experiment are depicted in Fig.~\ref{fig2}.

\textit{Readout:-} In order to perform a dispersive readout of  the qubit, we tune all qubit to their idle frequencies quickly at time $t_0+t$. This enables us to maintain the final state after the evolution to make the projective measurements to obtain $\langle\hat{a}_l^{\dagger}\hat{a}_l(t)\rangle$.

\section{Calibrate all qubits to their working frequencies}
In our experiment, we need to apply pulse signals to detune the qubits from their idle points to the working frequencies. In addition to this, we further need to exploit the Z line control to induce an oscillation around the working frequencies. However, due to the crosstalk, inaccurate correspondence of pulse amplitude and qubit frequency, it is experimentally challenging to achieve the desired working frequency profiles. Therefore,  we have to perform a calibration to reduce the deviation from the desired frequency profile, as we describe below.

As a first step, we correct the pulse distortion and pulse cross talk to get standard pulses that are independent of each other~\cite{Yan2019}. This procedure enables us to avoid the effect of pulse deformation and other undesired effects. Although this procedure can significantly reduce the deviation from the desired frequency setup, the deviation can be large when the working frequencies depend on space, thus we need further calibrations.

\begin{figure}
	\centering
	\includegraphics[width=0.97\textwidth]{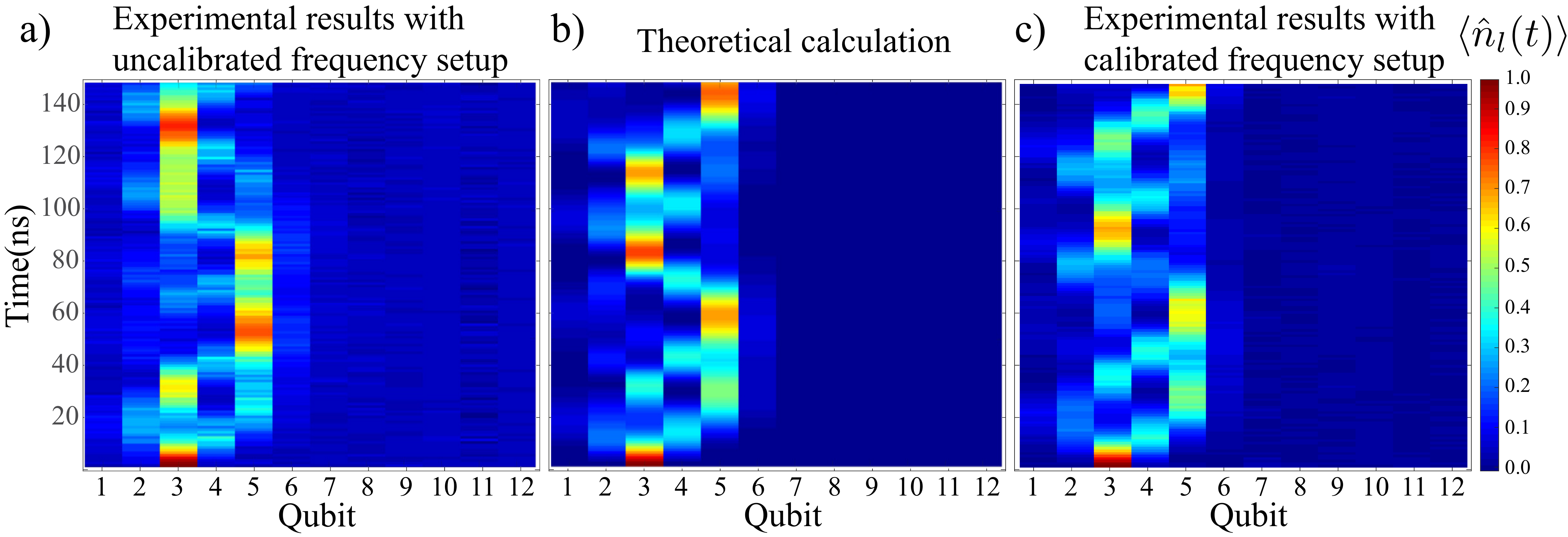}
	\caption{
		Fourth stage of the calibration procedure. a) Depicts the experimental results before the 4th stage of calibration and b) the theoretical results calculated using the designed frequency setup. There is  a clear deviation between the theory and experiment. To fix this problem, we theoretically find the actual frequency setup of the array. By using the Z control lines we apply a pulse offset in such a way that the theory and experiment match, as one can see from figures b) and c), respectively. 
	}
	\label{fig3}
\end{figure}

Secondly, we perform qubit-qubit swapping between two nearest-neighboring qubits to calibrate their resonant frequency. To do this, we detune the pair of neighboring qubits to their working frequencies, leaving other qubits at their idle  frequencies. Then we compare it with desired frequency spacing and add the 2nd offset pulse for calibration. In despite of this, there is a residual cross talk from the other qubits that creates a deviation of the qubit from the designed frequency. 

As a third calibration step, we perform an in-situ cross talk calibration~\cite{Yan2019}. The target qubit to be calibrated and its two nearest-neighbor qubits are detuned to their corresponding working frequencies. Meanwhile, during the calibration we keep the rest of the qubits at their idle frequencies. Next, we apply pulses to the neighbors of the target qubit. Those pulses have an amplitude 0.2 times smaller than the usual one. This procedure shows us the actual frequency of the target qubit. Subsequently, we calculate the 3rd offset pulse amplitude through the frequency difference between actual frequency and designed frequency. 

In the fourth step of the calibration, we resort on the theoretical model. In order to isolate the possible sources of errors, we perform a calibration in the absence of modulation pulses.
As shown in Fig.~\ref{fig3}, we perform experimental measurements of the populations $\langle\hat{n}_l(t)\rangle$ without the time dependent modulation, where $\hat{n}=\hat{a}_l^{\dagger}\hat{a}_l$, by using the static profile $g_l$ of working frequencies. Subsequently, we compare the dynamics of theoretical result $\langle\hat{n}_l(t)\rangle_{\text{theory}}$ calculated with the expected working frequencies with the experimental result $\langle\hat{n}_l(t)\rangle_{\text{exp}}$. Due to experimental errors, the theoretical result shows a deviation of few MHz from the experimental data. In order to find the corrected working frequency setup, we theoretically explore the parameter space to find the frequency profile that matches the experimental result. In this way, we think we truly get the actual working frequency setup of the experiment, through which the 4th pulse offset can be calculated.

After these four calibration steps, the experimental dynamics of the populations without frequency time modulation is almost the same as the predicted results, as shown in figure~\ref{fig3}. This is a good indicator that the experimental working frequency setup is very close to the designed working frequency setup.
Afterwards, we investigate the effect of time dependent modulation on the frequency profile of the qubits. With this aim, we use the same method as described above: We experimentally measure the dynamics of the populations and compare it with the expected theoretical results. Due to the time modulation, further errors appear, and we need to theoretically fit the amplitudes of the frequency modulation in order to match the measured population dynamics with uncalibrated pulse. With these results at hand, we can find the offset amplitude to be compensated. However, 
the compensation seems to have little effect on the evolution. We speculate that the origin of the deviation from theory is due to additional cross talk of pulse signals with high frequency,  and not just to the individual signal amplitude.

\section{ZZ Correlation Measurement}
Our experimental setup give us access to the the full correlation matrix of the photons. In particular, we focus here on the correlation function $C_{ZZ}(i,j,t)=\langle\sigma^z_{i}(t)\sigma^z_{j}(t)\rangle-\langle\sigma^z_{i}(t)\rangle\langle\sigma^z_{j}(t)\rangle$ with $\sigma^z_{i}=2\hat{n}_{i}-1$ reveals signatures of the proximity effect.
To measure the ZZ correlation of $C_{ZZ}(i,j)$, we jointly readout the states of twelve qubits and get the population distributions of the 12-qubit states $|Q_1...Q_{12}\rangle$. Then we measure the probability $P_{00}(i,j)$ that both qubit $i$ and qubit $j$ are at the state zero, the probability $P_{01}(i,j)$ that qubit $i$ is at the state zero and qubit $j$ is at state one, the probability $P_{10}(i,j)$ that qubit $i$ is at the state one and qubit $j$ is at state zero, and the probability $P_{11}(i,j)$ that both qubit $i$ and qubit $j$ are at the state one. We also measure the probabilities $P_0(i)$ and $P_1(i)$ of the $i-$th qubit to be at state $0$ and $1$, respectively. Similarly, we measure the probabilities $P_0(j)$ and $P_1(j)$ for the $j-$th qubit.
Once we have all this information, we can obtain the ZZ Correlation matrix $ \sigma^z_{i}\sigma^z_{j}$, which is given by
$$
C_{ZZ}(i,j) = P_{00}(i,j)+P_{11}(i,j)-P_{01}(i,j)-P_{10}(i,j)-[P_0(i)-P_1(i)][P_0(j)-P_1(j)] .
$$

\section{Finite-size signatures of parametric resonance in the group of qubits $Q_1-Q_{6}$}
In this section of the supplemental information, we discuss in detail finite-size signatures of parametric resonance in our experiment. For this purpose, we theoretically obtain the resonance condition that motivate our choice for the driving frequency $\omega/2\pi=19.67$~MHz that we use in the experiment.

\begin{figure}
	\centering
	\includegraphics[width=0.97\textwidth]{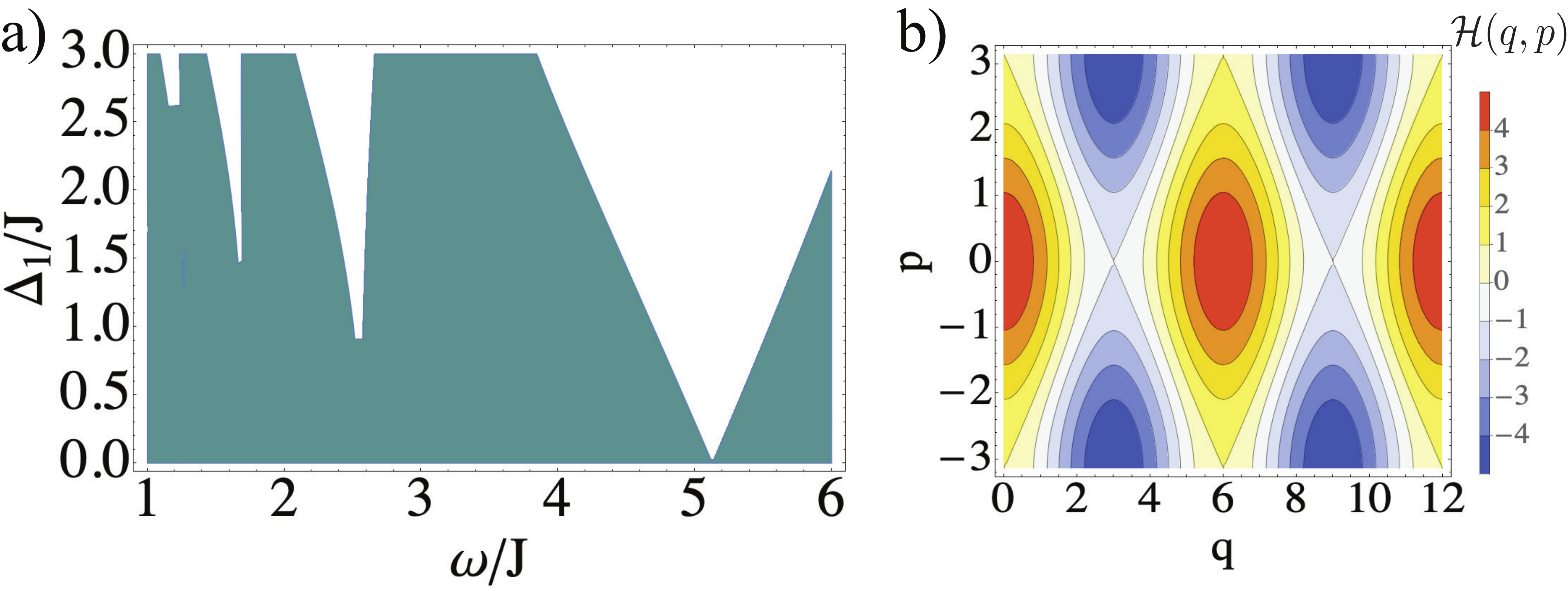}
	\caption{Parametric resonance in the semiclassical model. a) The stability diagram of the fixed point $(q,p)=(L,0)$ as a function of the driving frequency $\omega$ and the amplitude of the drive $g_1$. The coloured regions represent the set parameters for which the fixed point $(q,p)=(L,0)$ is stable. Contrary to this, for parameters in the white regions, the system is on parametric resonance and the fixed point becomes unstable. b) Shows the level contours of the semiclassical potential of the lattice in the absence of driving and disorder. In the absence of driving and disorder, the semiclassical potential associated to the lattice reads $\mathcal{H}(q,p,t)=\hbar\Delta_0\cos\left(\frac{2\pi q}{L}\right)+2\hbar J \cos\left(\frac{bp}{\hbar}\right)$, in the case of a cosine potential. The linearization procedure gives us information about the stability of the red regions in the contour plot. When the driving is strong and the system is on parametric resonance, it exhibits chaotic dynamics.
	}
	\label{fig4}
\end{figure}

In our experiment, the qubits $Q_1-Q_{6}$ are used to simulate the ergodic domain. However, in order to derive a semiclassical model, it is convenient to consider the Hamiltonian in the limit $N\gg1$

\begin{align}
\label{eq:HamiltonianErgodicPBCSI}
\hat{H}_{\text{erg}}(t)&=\hbar\sum^{N/2}_{l=1}g_{l}(t) \hat{n}_{l}+ \hbar J\sum^{N/2-1}_{l=1}(\hat{a}_{l}\hat{a}^{\dagger}_{l+1}+\text{h.c}) \ ,
 \end{align}
where $g_{l}(t)=\bar{g}+[\Delta_0+\Delta_1\cos(\omega t)]\cos(4\pi l/N)$.
In our experiment, we work with a finite system with $N=12$ lattice sites. However, as we show here, we can obtain relevant information about the system by resorting to a semiclassical description. To derive the semiclassical model, we
we assume periodic boundary conditions $\hat{a}_{l+N}=\hat{a}_N$ and define a Fourier transformation $\hat{a}_j=\frac{1}{\sqrt{L}}\sum_k e^{\mathrm{i}k b j}A_k$, where the condition $kL=2n\pi$, with integer $n$, must be satisfied. The parameter $b$ denotes the lattice constant and $L=bN/2$ is the length of the chain. In the continuum limit $L\gg 1$, the quasimomentum becomes a continuum variable and the Hamiltonian Eq.\eqref{eq:HamiltonianErgodicPBCSI} in the reciprocal space can be replaced by an integral over the momentum $-\hbar\pi/b < p<\hbar\pi/b$, where $p=\hbar k$, as follows
\begin{equation}
         \label{eq:ReciprocalSpaceHam}
\hat{H}_{\text{erg}}=\int^{\hbar\pi/b}_{-\hbar\pi/b}\Psi^{\dagger}_{p}\hat{\mathcal{H}}_{\text{erg}}(\hat{q},\hat{p},t)\ \boldsymbol{\tau^z}_p\Psi_{p}\ dp
\ ,
\end{equation}
 where $\Psi^{\dagger}_{p}=(A_{p}^{\dagger}, A_{-p})$  and $\hat{q}=\mathrm{i}\hbar\partial_p$ is the position operator in reciprocal space. For convenience, we have introduced the Pauli matrix $\boldsymbol{\tau^z}_p=\text{diag}(1,-1)$ in the particle-hole basis. The Hamiltonian $\hat{\mathcal{H}}_{\text{erg}}(\hat{q},\hat{p},t)$ represents is the Hamiltonian in first quantization. In addition, if we define the operators $\hat{Q}=2\pi \hat{q}/L$ and $\hat{P}=b\hat{p}/\hbar$, we can show that they satisfy the commutation relation $[\hat{Q},\hat{P}]=2\pi\hbar\mathrm{i}/N$, because $[\hat{q},\hat{p}]=\mathrm{i}\hbar$. In the continuum limit $N\gg1$, the operators $\hat{Q}$ and $\hat{P}$ behave like classical variables because their commutation relation scales as $1/N$. Note that from here we can extract the classical Hamiltonian[]
\begin{align}
\label{eq:HamiltonianClassicalErgodicSI}
\mathcal{H}_{\text{erg}}(q,p,t)=\hbar[\Delta_0+\Delta_1\cos(\omega t)]\cos\left(\frac{2\pi q}{L}\right)+2\hbar J \cos\left(\frac{bp}{\hbar}\right)\ 
 \end{align}
 in a rotating frame with an angular frequency $\bar{g}$. The canonical coordinates $q$ and $p$ are related to the mean position and momentum of the excitation, respectively. In the absence of drive ($\Delta_1=0$) and in despite of the nonlinear character of the Hamiltonian~\eqref{eq:HamiltonianClassicalErgodicSI}, the energy is conserved and the system exhibits regular motion.

 In order to gain some intuition about parametric resonance, we linearize the Hamiltonian around the fixed point $(q,p)=(L,0)$, as follows
\begin{align}
\label{eq:HamiltonianClassicalErgodicLinearSI}
\mathcal{H}_{\text{erg}}(\delta q,\delta p,t)=\hbar[\Delta_0+\Delta_1\cos(\omega t)]\left[1-\frac{1}{2}\left(\frac{2\pi }{L}\right)^2 (\delta q)^2\right]+2\hbar J \left[1-\frac{1}{2}\left(\frac{b}{\hbar}\right)^2(\delta p)^2\right]
\ ,
 \end{align}
where $\delta q$ and $\delta p$ are small deviations of the canonical coordinates about the fixed point. Interestingly, the Hamiltonian~\ref{eq:HamiltonianClassicalErgodicLinearSI} has the same form as a parametric oscillator, where the angular frequency of small oscillation is given by $\Omega=4\pi\sqrt{2\Delta_0 J}/N$. If we now drive the system with an angular frequency $\omega$ that is conmensurable with $\Omega$, the system exhibits parametric resonance and the fixed point becomes unstable. In the experimental setup, we tuned the drive such that the resonance condition $m\omega=2\Omega$ with $m=3$. Furthermore, as we are working in the parameter regime $\Delta_0=\Delta_1=3J$ with $J/2\pi=11.5$~MHz, we need to drive with a frequency $\omega/2\pi=19.67$~MHz to achieve finite size signatures of parametric resonance.

 \section{Floquet theory and quasienergy level statistics}

In our experiment, we simulate a time periodic Hamiltonian, that is, $\hat{H}(t+T)=\hat{H}(t)$, where $T=2\pi/\omega$ is the period of the drive.
Due to the time-periodic character of the Hamiltonian, Floquet theory is a powerful tool to describe the dynamics of the system~\cite{Hanggi1991,Haenggi1998,Holthaus2005}. With purpose, we use the Floquet operator $\hat{\mathcal{F}}=\hat{U}(T)$, which is the evolution operator $\hat{U}(t)$ within one period of the drive. The most important information can be obtained by solving the eigenvalue problem $\hat{\mathcal{F}}|\Phi_{\alpha}\rangle=e^{-\mathrm{i}\varepsilon_{\alpha}T/\hbar}|\Phi_{\alpha}\rangle$. The eigenvectors $|\Phi_{\alpha}\rangle$ are known as the Floquet states and $-\hbar\omega/2\leq\varepsilon_{\alpha}\leq\hbar\omega/2$ are the quasienergies~\cite{Hanggi1991,Haenggi1998,Holthaus2005}.  

As a diagnosis of localization, we can resort on the statistics of the ratios $r_{\alpha}=\min(\delta_{\alpha},\delta_{\alpha+1})/\max(\delta_{\alpha},\delta_{\alpha+1})$ with $\delta_{\alpha}=\varepsilon_{\alpha+1}-\varepsilon_{\alpha}$ as in Ref.~\cite{Rigol2014}. When the system is localized in space, the quasienergies are uncorrelated and the statistics of the gaps follows a Poissonian distribution $P_{\text{Poisson}}(r)=2/(1+r)^2$. Contrary to this, when the system is in the ergodic phase, there is strong level respulsion and the quasienergy levels become highly correlated~\cite{roushan17, Rigol2014}. As a consequence, the spectrum of the Floquet operator exhibits the same statistics as the circular ortogonal ensemble (COE) of random matrices. In this case, the statistics of ratios follows the distribution~\cite{Rigol2014,tangpanitanon2019quantum}
\begin{align}
 P_{\text{COE}}(r)&=\frac{2}{3}\left\{\left[\frac{\sin\left(\frac{2\pi r}{r+1}\right)}{2\pi r^2}\right]+\frac{1}{(1+r)^2}+\left[\frac{\sin\left(\frac{2\pi}{r+1}\right)}{2\pi }\right]\right\}
 \nonumber\\&
 -\frac{2}{3}\left\{\left[\frac{\cos\left(\frac{2\pi }{r+1}\right)}{2\pi r^2}\right]+\left[\frac{\cos\left(\frac{2\pi r}{r+1}\right)}{r(r+1)}\right]\right\}
 \ .
\end{align}
In figure 1~e) of the main text, we depict the numerical results for the level statistics in order to unveil signatures of the proximity effect~\cite{Bastidas2018}.

\clearpage
\newpage


%
